\documentstyle[psfig]{l-aa}

\begin{document}

   \thesaurus{07          
                           editors
              (08.01.1;   
               08.05.3;   
               08.16.4;   
               08.09.2: Sakurai's object;   
               08.22.3)}   

   \title{
The rapid evolution of the born-again giant Sakurai's object
}

   \author{M. Asplund\inst{1}, 
D.L. Lambert\inst{2}, T. Kipper\inst{3}, D. Pollacco\inst{4} and M.D. Shetrone\inst{5} }

   \offprints{M. Asplund (martin@nordita.dk)}

   \institute{              
              NORDITA, 
              Blegdamsvej 17, 
              DK-2100 ~Copenhagen {\O}, 
              Denmark\\          
              \and
              Department of Astronomy,
              University of Texas,
              Austin, TX 78712,
              USA \\
              \and
              Tartu Observatory,
              T{\~o}ravere,
              EE2444,
              Estonia\\
              \and
	      Isaac Newton Group, 
	      Apartado de Correos 368,
	      Santa Cruz de La Palma, 
	      Tenerife 38780, 
	      Spain\\
              \and
              McDonald Observatory,
              P.O. Box 1337,
              Fort Davis, Tx 79734,
              USA \\
              }

   \date{
Received: Sept. 7, 1998; accepted: Sept. 30, 1998 
}

   \maketitle

   \markboth{Asplund et al.:
The rapid evolution of the born-again giant Sakurai's object 
}
{Asplund et al.: 
The rapid evolution of the born-again giant Sakurai's object 
}

\begin{abstract}  

The extraordinarily rapid evolution of the born-again giant Sakurai's
object following discovery in 1996 has been investigated. 
%
The evolution can be traced both in a continued
cooling of the stellar surface and dramatic changes in chemical
composition on a timescale of a mere few months.
The abundance alterations are the results of
the mixing and nuclear reactions which have ensued due to
the final He-shell flash which occurred during the 
descent along the white dwarf cooling track. 
The observed changes in the H and Li abundances
can be explained by ingestion and burning of the
H-rich envelope and Li-production
through the Cameron-Fowler mechanism.
The rapidly increasing abundances of the light $s$-elements
(including Sc) is consistent with current $s$-processing  
by neutrons released from
the concomitantly produced $^{13}$C. 
However, the possibility that the $s$-elements have previously
been synthesized during the AGB-phase 
and only mixed to the surface in connection with the
final He-shell flash in the pre-white dwarf cannot be convincingly
ruled out either.
%
Since Sakurai's object shows substantial abundance similarities with the
R\,CrB stars and has recently undergone R\,CrB-like visual
fading events, the ``birth'' of an R\,CrB
star may have been witnessed for the first time ever.
Sakurai's object thus lends strong support for the suggestion that
at least some of
the R\,CrB stars have been formed through a final He-shell flash
in a post-AGB star. 


\end{abstract}

\section{Introduction}

Normally stellar evolution proceeds on time-scales of millions
or billions of years.
Notable exceptions are e.g. supernovae, 
though such cataclysmic
events only mark the end point of the evolution as normal stars.
Witnessing stellar evolution in real time
without a simultaneous complete destruction of the star
is thus very rare. 
``Born-again'' giants do, however, offer
such a remarkable opportunity, which offers important
windows through which stellar nucleosynthesis and evolution
can be glimpsed.

On its initial descent of the white dwarf cooling track, a star
may be ballooned back to supergiant dimensions  
by a final He-shell flash which is triggered by the 
compressional heating of the stellar interior.
Rather than fading steadily to oblivion as a white dwarf the star 
makes a second appearance as a luminous giant, a ``born-again'' giant.
Theoretical estimates suggest that about 10\% of all low- and
intermediate mass stars going through a planetary nebula phase
will  experience such a final He-shell flash 
(Renzini 1990; Iben et al. 1996).
The observational rarity of such events is the result of
the very short life-time as a born-again giant -- typically 
100-1000 years, depending mainly on the stellar envelope mass -- 
before it completes its loop in the 
Hertzsprung-Russell diagram and once again starts contracting
to become a white dwarf, this time for good.

To date, only three stars are believed to have been observed
going through a born-again phase: V605\,Aql (Nova Aql 1919), 
FG\,Sge and the
recently discovered Sakurai's object (V4334\,Sgr)
(Nakano et al. 1996).
A further handful of stars (e.g. Abell 30 and Abell 78) 
presently in a seemingly second 
stage as planetary nebulae may be identified as having 
recently been born-again giants. 
Furthermore, the R\,CrB variables could also possibly be born-again
giants (Asplund et al. 1997b; Asplund et al. 1998;
Lambert et al. 1998); 
their H-deficiency can be explained if the final He-shell
flash occurs after the outer H-burning shell has been
extinguished, in which
case the convection zone due to the He-flash may ingest and burn the
H-rich envelope.

Both V605\,Aql, FG\,Sge and Sakurai's object have shown a remarkably fast
stellar evolution.
From having been an O star at the end of last century
(van Genderen \& Gautschy 1995),
FG\,Sge has continued to brighten in the visual while cooling,
presumably the result of the expansion following the final He-shell
flash (see Kipper 1996 for a recent review).
Furthermore, an enrichment of $s$-process elements
which took place within seven years has been reported
(Langer et al. 1974). Lately, FG\,Sge has also started to show
R\,CrB-like variability (Papousek 1992), 
which may classify the star as
a ``new-born'' R\,CrB star, in particular since there are 
hints that the star may be H-deficient (Gonzalez et al. 1998).
After discovery in 1996, Sakurai's object has shown an
exceedingly rapid evolution, even out-pacing 
FG\,Sge; whether or not it is evolving faster than V605\,Aql
is difficult to judge considering the scarceness of data
from the outburst in 1919 but the evolutionary speed for
V605\,Aql may have been comparable (Clayton \& de Marco 1997). 
Like FG\,Sge Sakurai's object has cooled significantly
(Asplund et al. 1997b; Duerbeck et al. 1997).
More impressively though, its chemical
composition seems to have been altered drastically on a time-scale 
of a mere few months (Asplund et al. 1997b). 
FG\,Sge's title as the ``fastest evolving star ever identified''
(van Genderen \& Gautschy 1995)
is certainly called in doubt by Sakurai's object.

Sakurai's object offers a unique opportunity to 
study  stellar evolution and attendant nucleosynthesis in real time.
Several papers has been devoted
to the star and its remarkable evolution 
(e.g. Asplund et al. 1997b; Duerbeck \& Benetti 1996;
Duerbeck et al. 1997; Eyres et al. 1998; 
Jacoby et al. 1998; Kimeswenger \& Kerber 1998; 
Kipper \& Klochkova 1997; Shetrone \& Keane 1998).
However, the  published abundance analyses of the star have
relied on very different model atmospheres and spectral features,
making a direct comparison of the results difficult.
Therefore we have re-analysed the available
published (Asplund et al. 1997b; Kipper \& Klochkova 1997;
Shetrone \& Keane 1998) and unpublished (D. Pollacco and G. Wallerstein)
spectra in a homogeneous way, using the same model 
atmospheres and whenever possible the same spectral lines to
avoid as far as possible systematic differences.
The resultant map of the evolution of the composition offers
important clues to  the star's immediate past history.

\section{Observations}

Some of the observations on which the present study is based have
previously been used by Asplund et al. (1997b), 
Kipper \& Klochkova (1997) and Shetrone \& Keane (1998) for
abundance analyses. For completeness we repeat
the observational details here, together with the relevant information on the
previously unpublished spectra.

The spectra of Asplund et al. (1997b) cover 3\,700-10\,150\,\AA $ $ 
at a resolving power of about 30\,000 and 
were obtained with 
the 2.7\,m telescope at McDonald Observatory on May 5 and 6 
and on October 7,  1996. 
Another spectrum was acquired with the McDonald 2.1\,m telescope
on May 9, 1996, which covers the region
5\,720\,\AA\ to 7\,200\,\AA\ at a resolving power of about 60\,000;
the three May spectra have been combined to a single line list. 
Shetrone \& Keane (1998) observed Sakurai's object on April 20,
1996, with CTIO 4\,m telescope with a resolving power of about 14\,000.
The spectrum covers the region 3\,800-7\,500\,\AA $ $ with S/N
ranging from 100 to 200, depending on the wavelength region.
In addition G. Wallerstein (unpublished research) obtained a spectrum
of the star with the CTIO 4\,m telescope at a resolving power of 
about 12\,500 on April 25.
The two April spectra have here been combined to one line list
though greater weight has been given to the April 20 spectrum due
to its higher quality. 
A spectrum obtained with the William Herschel 4.2\,m telescope 
(Pollacco, unpublished research) was acquired on June 4, 1996, at a resolving
power of 45\,000 and S/N of about 70. The spectrum covers the
wavelength region 3\,800-8\,000\,\AA .
Finally, Kipper \& Klochkova (1997) obtained spectra of the
star on July 3, 1996, with the echelle spectrometer on the 
Russian 6\,m telescope,
providing a spectral resolving power of about 20\,000 and S/N of around
75. The wavelength coverage was 5\,000-8\,000\,\AA .
In addition, other spectra obtained by us or kindly provided
by various colleagues have been attempted for abundance analyses,  but
in all cases the resolution proved insufficient for reliable results, 
and they are therefore not included in the present study. 

With the exceptions of a few spectral regions for which
spectral synthesis was carried out (see below),
the abundance analysis is based on lines for which
equivalent widths could be measured readily. The line lists
for the April, June and July spectra were modelled following the
lines used by Asplund et al. (1997b) for the May and October 
observations whenever the spectral coverage allowed. The
lines are thus in general not the same in the present study 
as in the original works by Shetrone \& Keane (1998) and
Kipper \& Klochkova (1997).
Spectral synthesis was applied to a few wavelength regions 
of particular interest: 
H$\alpha$, H$\beta$, the C$_2$ Swan 0-1 and 1-0 band-heads, 
the Li\,D doublet at 6\,707\,\AA,
the blended C\,{\sc ii} lines at 
6\,578.05\,\AA $ $ and 6\,582.88\,\AA $ $ and the He {\sc i} D$_3$
line at 5\,876\,\AA , which is blended by two 
C\,{\sc i} lines.
The adopted atomic data were the same as in Lambert et al. (1998). 
The $gf$-values for the C\,{\sc i} lines present in the above
synthesized regions were decreased according to the
magnitude of the ``carbon problem'' (typically 0.7\,dex)
of the individual spectra, see Sect. \ref{s:param}. 

\section{Derivation of the stellar parameters
\label{s:param}}

\begin{figure}[t]
\centerline{
\psfig{figure=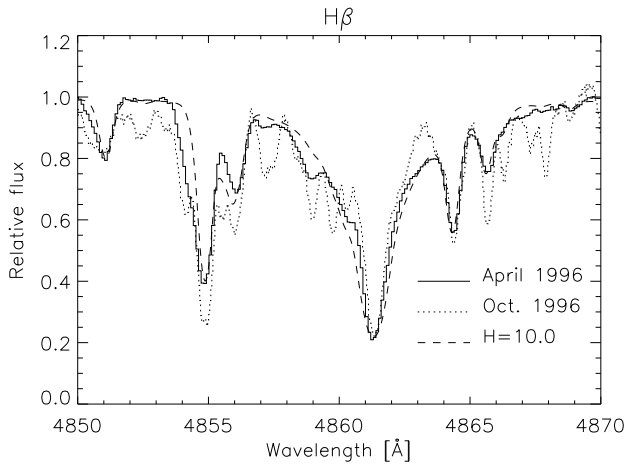,width=9.cm}}
\caption{H$\beta$ in Sakurai's object in April (solid) compared
with in October (dotted) 1996. The dashed line represents the
spectral synthesis of the April spectrum using a H-abundance of
10.0. Note that all features but H$\beta$ had became stronger 
in October as a result of the decreasing continuous opacity
with a diminishing H-content and in some cases an increased
elemental abundance (e.g. the $s$-elements)
}
         \label{f:hbeta}
\end{figure}

In order to
derive accurate stellar parameters a range of different $T_{\rm eff}$
and log\,$g$ criteria was utilized: ionization balance 
(mainly N\,{\sc i}/N\,{\sc ii},
Si\,{\sc i}/Si\,{\sc ii}, Cr\,{\sc i}/Cr\,{\sc ii}, 
and Fe\,{\sc i}/Fe\,{\sc ii}),
excitation balance (O\,{\sc i}, Fe\,{\sc i} and Fe\,{\sc ii}),
and line strengths of sensitive spectral features 
(C$_2$, C\,{\sc ii} and He\,{\sc i} lines). 
The extended line wings of H$\alpha$ and H$\beta$ allowed a determination
of the stellar H abundance,
but the profiles also provided
additional information on the adopted $T_{\rm eff}$
and log\,$g$ (Fig. \ref{f:hbeta}).
The microturbulence parameter $\xi_{\rm t}$
was estimated mainly from 
C\,{\sc i}, Ti\,{\sc ii}, Fe\,{\sc i}, Fe\,{\sc ii} and Y\,{\sc ii}
lines of different strengths; to within the uncertainties all
species indicated the same $\xi_{\rm t}$.
The derived temperatures from $B-V$ colours (Duerbeck et al. 1997)
assuming a reddening of $E(B-V) = 0.70$ (N.K. Rao, private
communication) provided another temperature estimate which
was in very good agreement with the other criteria
(Fig. \ref{f:sakteff}). 
Line-blanketed, hydrogen-deficient model atmospheres similar to those
described by Asplund et al. (1997a) but with a range of hydrogen
abundances especially constructed for the present analysis
formed the basis of the investigation.
  
Since the present study of the May and October 1996 spectra 
is based on the same observations, spectral lines and model
atmospheres as in Asplund et al. (1997b),
the previously derived stellar parameters for May 
($T_{\rm eff} = 7\,500\pm 300$\,K, 
log\,$g = 0.0\pm 0.3$, $\xi_{\rm t} = 8.0\pm 1.0$\,km\,s$^{-1}$) 
and October ($T_{\rm eff} = 6\,900\pm 300$\,K, log\,$g = 0.5\pm0.3$,
$\xi_{\rm t} = 6.5\pm1.0$\,km\,s$^{-1}$) did not warrant a 
re-consideration.
Within the framework of the present analysis the April spectrum is
characterized by $T_{\rm eff} = 7\,750\pm 300$\,K, 
log\,$g = 0.25\pm 0.3$ and $\xi_{\rm t} = 10.5\pm 1.0$\,km\,s$^{-1}$. 
The June spectrum indicates $T_{\rm eff} = 7\,400\pm 300$\,K, 
log\,$g = 0.4\pm 0.3$ and $\xi_{\rm t} = 8.5\pm 1.0$\,km\,s$^{-1}$. 
Similarly the July spectrum is adequately described using
$T_{\rm eff} = 7\,250\pm 300$\,K, 
log\,$g = 0.5\pm 0.3$ and $\xi_{\rm t} = 8.5\pm 1.0$\,km\,s$^{-1}$.
The slight differences compared with the original analyses of 
Kipper \& Klochkova (1997) and Shetrone \& Keane (1998) are not
alarming considering the very different adopted model atmospheres as
well as the use of different $T_{\rm eff}$-log\,$g$ criteria.
The quoted uncertainties in the derived parameters represent the
discrepancies between the various indicators; additional systematic
errors are not accounted for. However, since the same criteria has been
used, the systematic errors are likely very similar for the different
observing epochs, and thus will not compromise our conclusions 
regarding the very rapid evolution of the star.

\begin{table}[t!]
\caption{The estimated stellar parameters for Sakurai's object
in 1996
\label{t:param}
}
\begin{tabular}{lcccc} 
 \hline
Epoch       & $T_{\rm eff}$   & log\,$g$     & $\xi_{\rm t}$ & C/He \\ 
            &  [K]            &  [cgs]       & [km s$^{-1}$] &      \\             
\hline \\
April 20-25 & $7\,750\pm300$  & $0.25\pm0.3$ & $10.5\pm1.0$  & 10\% \\
May 5-9     & $7\,500\pm300$  & $0.00\pm0.3$ & $8.0\pm1.0$   & 10\% \\  
June 4      & $7\,400\pm300$  & $0.40\pm0.3$ & $8.5\pm1.0$   & 10\% \\  
July 3      & $7\,250\pm300$  & $0.50\pm0.3$ & $8.5\pm1.0$   & 10\% \\  
October 7   & $6\,900\pm300$  & $0.50\pm0.3$ & $6.5\pm1.0$   & 10\% \\ 
\hline
\end{tabular}

\end{table}
   
For all model atmospheres except for a few test cases, 
a C/He ratio (by number) of 10\% has been adopted, following 
comparison of the C\,{\sc ii} and He\,{\sc i} line strengths in the
April and May spectra. It should be emphasized, however, that these lines do not
provide a very accurate estimate of the C/He ratio in cool H-deficient
stars such as Sakurai's object due to the 
sensitivity of the lines to the adopted stellar parameters and the
significant contribution of blending C\,{\sc i} lines with their 
unexplained "carbon problem" (see below) to the 
He\,{\sc i} 5\,876\,\AA$ $ triplet.
The June and July spectra are consistent with the adopted C/He ratio 
of 10\%, though a better agreement between the 
C\,{\sc ii} and He\,{\sc i} lines would be obtained with a
slightly higher ratio. The $T_{\rm eff}$ in October was too low
to enable a determination of the C/He ratio through the
high-excitation He\,{\sc i} and C\,{\sc ii} lines.

In the H-deficient and C-rich photospheres of Sakurai's object
and the R\,CrB stars, the continuous opacity at the flux-carrying
wavelengths is no longer provided by hydrogen but by photoionization
of highly excited levels of C\,{\sc i}. Furthermore, carbon is by
far the most important electron donor in the line-forming regions.
The atmospheric
structure is therefore essentially completely determined by C
(Asplund et al. 1997a).
Since the photospheric C abundance is derived from lines
originating from levels with
only slightly lower excitation energy
than those providing the continuous opacity, 
the strengths of C\,{\sc i} lines in these stars should be 
essentially independent
on both the C abundance and the adopted $T_{\rm eff}$ and log\,$g$,
which is also confirmed by observations and
theoretical calculations (Lambert et al. 1998). 
Therefore, the theoretical line strengths of weak C\,{\sc i} lines 
should agree with observations since there are no 
free parameters to tune in order to achieve agreement 
as in normal analyses. This is unfortunately
neither the case for the R\,CrB stars 
(Gustafsson \& Asplund 1996; Asplund et al. 1997a;
Lambert et al. 1998) nor for Sakurai's object  
(Asplund et al. 1997b; Kipper \& Klochkova 1997);
all analyses reveal a significant discrepancy between the predicted
and observed line strengths to the extent that the derived abundance
is about 0.6\,dex smaller than the input C abundance for the model
atmospheres. 
The explanation for this ``carbon problem'' is still unknown, but it may
be related to inappropriate assumptions for the construction
of the model atmospheres (1-D, static, flux-constant atmospheres
in LTE) (Lambert et al. 1998). 

Sakurai's object also shows the same discrepancy by 
on average 0.7\,dex.
As will be illustrated further below, the abundances of most
elements have remained the same throughout 1996, which suggests
that the C/He ratio has not changed; an increasing C
abundance would manifest itself as a decrease in derived
abundances for all elements when assuming a constant C/He in
the analysis
unless for some reason X/C remained the same for all
elements except for He.
Fortunately, various tests (Lambert et al. 1998) suggest that
although absolute abundances can be severely affected, 
abundance ratios such as [X/Fe] are largely immune to the carbon 
problem because of the similar sensitivity to the atmospheric 
structure of most elements;
changing C/He from 10\% to 1\% introduces differences in 
[X/Fe] by $\la 0.1$\,dex (Asplund et al. 1998).
Therefore, conclusions regarding the evolutionary history of
Sakurai's object and the R\,CrB stars can still be drawn,
in spite of the unsolved dilemma caused by the carbon problem.

\section{Chemical composition}

\begin{table*}[t]
\caption{ Chemical compositions of Sakurai's object, the 
R\,CrB majority stars, the R\,CrB star V854\,Cen and the Sun
(normalized to log\,($\Sigma \mu_i \epsilon_i$) = 12.15).
The errors quoted in the table are the  
standard deviations for the different lines
of a species; in case the abundance is derived from a single
line, no error is given. More uncertain values are marked with :
\label{t:abund}
}
\begin{tabular}{lcccccccc} 
 \hline
Element  & Sun$^{\rm a}$ & \multicolumn {5} {c} 
  {Sakurai's object in 1996} & 
R\,CrB  & V854\,Cen$^{\rm c}$ \\ 
 && April 20-25 & May 5-9 & June 4 & July 3 & October 7 & majority$^{\rm b}$  &  \\             
\hline \\
H  & 12.00 & 10.0~~~~~~~\, &  9.7~~~~~~~\,& 9.7~~~~~~~\,& 9.6~~~~~~~\,   
   & 9.0~~~~~~~\,&  $<4.1 - 6.9$ & 9.9 \\
He & 10.93 & 11.4$^{\rm d}~~~~~~~$ & 11.4$^{\rm d}~~~~~~~$ 
   & 11.4$^{\rm d}~~~~~~~$ & 11.4$^{\rm d}~~~~~~~$ & 11.4$^{\rm d}~~~~~~~$ 
   & 11.5$^{\rm d}$ & 11.4$^{\rm d}$   \\
Li &  3.31 & 3.6~~~~~~~\, & 3.6~~~~~~~\, & 3.6~~~~~~~\, & 4.0~~~~~~~\,&  4.2~~~~~~~\, 
   & $<1.1 - 3.5$ & $<2.0$  \\ 
C  &  8.52 & 9.7$^{\rm e}\pm0.2$ & 9.7$^{\rm e}\pm0.2$ & 9.6$^{\rm e}\pm0.2$ 
   & $9.7^{\rm e}\pm0.3$ &  9.8$^{\rm e}\pm0.3$ & 8.9$^{\rm e}$ 
   & $9.6^{\rm e}$ \\
N  &  7.92 & $9.0\pm0.3$ &  $8.9\pm0.4$ & $9.0\pm0.4$ & $9.2\pm0.3$ &  $8.9\pm0.2$ &  8.6 & 7.8 \\  
O  &  8.83 & $9.2\pm0.2$ &  $9.5\pm0.3$ & $9.3\pm0.4$ & $9.3\pm0.1$ &  $9.4\pm0.2$ &  8.2 & 8.9 \\
Ne &  8.08 & $9.4\pm0.2$ &  $9.4\pm0.3$ & $9.5\pm0.3$ & $9.5\pm0.3$ &              &      &     \\
Na &  6.33 & $6.6\pm0.1$ &  $6.7\pm0.1$ & $6.5\pm0.1$ & $6.6\pm0.2$ &  $6.8\pm0.1$ &  6.1 & 6.4 \\
Mg &  7.58 & $6.5\pm0.4$ &  $6.6\pm0.4$ & $6.3:\pm0.4$ & $6.3:\pm0.4$&  $6.5\pm0.3$ &  6.4 & 6.2 \\
Al &  6.47 & $6.5\pm0.2$ &  $6.6\pm0.2$ & $6.5\pm0.3$ & $6.6\pm0.3$ &  6.3~~~~~~~\,&  6.0 & 5.7 \\  
Si &  7.55 & $7.3\pm0.0$ &  $7.1\pm0.2$ & $7.1\pm0.2$ & $7.1\pm0.0$ &  $7.5\pm0.2$ &  7.1 & 7.0 \\ 
P  &  5.45 & 6.2~~~~~~~\,&  $6.2\pm0.4$ & $6.1\pm0.4$& 6.3~~~~~~~\,&              &      &     \\
S  &  7.33 & $6.8\pm0.1$ &  $6.6\pm0.1$ & $6.5\pm0.2$ & $6.7\pm0.1$ &  $6.9\pm0.1$ &  6.9 & 6.4 \\          
K  &  5.12 &             &  $4.9\pm0.0$ & 4.7~~~~~~~\,& $5.0\pm0.1$ &  $5.0\pm0.0$ &      &     \\
Ca &  6.36 & $5.2\pm0.1$ &  $5.6\pm0.3$ & $5.4\pm0.3$ & $5.6\pm0.4$ &  $5.5\pm0.4$ &  5.4 & 5.1 \\  
Sc &  3.17 & $3.1\pm0.1$ &  $3.1\pm0.1$ & $3.2\pm0.1$ & $3.3\pm0.2$ &  $3.9\pm0.2$ &      & 2.9 \\
Ti &  5.02 & $4.0\pm0.1$ &  $4.1\pm0.2$ & $4.2\pm0.2$ & $4.4\pm0.2$ &  $4.6\pm0.2$ &      & 4.1 \\
Cr &  5.67 & $4.5\pm0.1$ &  $4.5\pm0.2$ & $4.7\pm0.2$ & $4.8\pm0.2$ &  $5.1\pm0.2$ &      & 4.2 \\
Fe &  7.50 & $6.4\pm0.2$ &  $6.4\pm0.2$ & $6.4\pm0.2$ & $6.6\pm0.2$ &  $6.6\pm0.3$ &  6.5 & 6.0 \\  
Ni &  6.25 & $6.1\pm0.3$ &  $6.1\pm0.4$ & $5.9\pm0.2$ & $6.0\pm0.2$ &  $6.2\pm0.2$ &  5.9 & 5.9 \\  
Cu &  4.21 &             &  $5.0\pm0.3$ & $5.0\pm0.2$ & $5.1\pm0.0$ &  $5.0\pm0.1$ &      &     \\
Zn &  4.60 & $4.9\pm0.2$ &  $4.8\pm0.2$ & 4.9~~~~~~~\,& 5.1~~~~~~~\,&  5.4~~~~~~~\,&  4.3 & 4.4 \\ 
Rb &  2.60 &             &  $<3.7$      & 4.2~~~~~~~\,&             &  4.6~~~~~~~\,&      &     \\
Sr &  2.97 & $4.7:\pm0.1$&  $4.9:\pm0.2$& $5.0:\pm0.4$&             &  $5.4:\pm0.0$&      & 2.2 \\
Y  &  2.24 & $3.2\pm0.3$ &  $3.3\pm0.3$ & $3.3\pm0.3$ & $3.7\pm0.2$ &  $4.2\pm0.2$ &  2.1 & 2.2 \\   
Zr &  2.60 & $3.0\pm0.2$ &  $3.0\pm0.3$ & $3.2\pm0.2$ & $3.3\pm0.2$ &  $3.5\pm0.3$ &      & 2.1 \\
Ba &  2.13 & $1.5\pm0.1$ &  $1.5\pm0.2$ & $1.5\pm0.2$ & $1.8\pm0.1$ &  $1.9\pm0.4$ &  1.6 & 1.3 \\  
La &  1.17 &             &  $<1.6$      &             & 1.3~~~~~~~\,&  1.5~~~~~~~\,&      & 0.4 \\
\hline
\end{tabular}

\begin{list}{}{}
\item[$^{\rm a}$] From Grevesse \& Sauval (1998). For Li the meteoritic 
value has been adopted.
\item[$^{\rm b}$] From Lambert et al. (1998). The abundances
are the mean from a sample of 14 stars; the four minority R\,CrB stars
are not included here.
\item[$^{\rm c}$] From Asplund et al. (1998).
\item[$^{\rm d}$] Input C/He ratio for model atmospheres: 
C/He=10\% estimated for Sakurai's object 
from the April and May spectra and V854\,Cen and 1\% assumed for 
R\,CrB stars.
\item[$^{\rm e}$] Spectroscopically determined C\,{\sc i} 
abundance, which differs from the input model atmosphere C abundance, 
see text.
\end{list}

\end{table*}
   
\subsection{Elemental abundances}
   
With the adopted stellar parameters described in Sect. \ref{s:param},
the resulting abundances under
the assumption of LTE as observed throughout 1996 are listed in
Table \ref{t:abund}.
To within the uncertainties the two April and the three May spectra
give identical abundances.
The abundances for May and October have been re-analysed for the 
present study, which has led to a few minor adjustments compared
with the original study (Asplund et al. 1997b): the abundances
of Ne, K, Fe, Cu and Zn have all increased by 0.1\,dex for May, as
a result of inclusion of additional lines, improved selection of
lines and adoption of better atomic data. 

From an inspection of Table \ref{t:abund} two important conclusions
are immediately obvious. First, the abundances of most elements 
(He, C, N, O, Ne, Na, Mg, Al, Si, P, S, K, Ca, Fe, Ni, Cu and La) 
have remained the same within the uncertainties
(about 0.3\,dex) throughout 1996. In most cases the total variation is
only about 0.2\,dex, which is remarkable considering the
intrinsic uncertainties in the analyses. Given the good agreement,
it is worthwhile emphasizing, 
that in all cases the adopted stellar parameters have been
determined before considering the absolute abundances and their 
implications, and no a posteriori fine-tuning has been allowed in order 
to improve the consistency. 
Second, several elements show abundance alterations which significantly
exceed the statistical scatter. This is true for
H (decrease by 1.0\,dex), Li ($+0.6$\,dex), Sc ($+0.8\,$dex),
Zn ($+0.7$\,dex), Rb ($> +0.9$\,dex) and Y ($+1.0\,$dex), and
probably also Ti ($+0.6$\,dex), Cr ($+0.6$\,dex), Sr ($+0.7$\,dex,
though uncertain due to only based on very strong lines) and 
Zr ($+0.5$\,dex). There is also some indication that the heavier
$s$-element Ba increased steadily in 1996,
though such a conclusion is not indisputable. 
Unfortunately, trustworthy abundances of other heavy
$s$-elements throughout 1996 which could verify such a finding
are lacking. 
The slightly larger variations (in total 0.4\,dex)
than the typical scatter for Si, S and
Ca are less likely to be signs of real abundance variations, since
no definite trend with time is apparent and only one of the
epochs show a disparate value.
We therefore attribute these variations to the uncertainties in 
the analyses.

For June and July the H abundance may be overestimated by 
$\sim 0.1-0.2$\,dex compared with the other epochs, since the
abundances are derived solely from H$\alpha$ as the spectra did not
cover H$\beta$. For the other times the greatest weight has been
given to H$\beta$, which in general needs slightly lower
abundance than H$\alpha$ for an acceptable fit to the extended line wings
in Sakurai's object.

\subsection{The reality of the abundance alterations}

Considering the  drastic abundance variations apparent 
in Sakurai's
object, it is natural to ask  
whether perhaps they can be blamed on erroneous  stellar
parameters or inappropriate approximations in the analysis.

It is important to realize that all abundance alterations cannot
be simultaneously annuled by {\em any} change in the adopted stellar 
parameters, regardless of how large the variation of the parameters.
The required $\Delta T_{\rm eff} \approx 1\,000$\,K to invalidate
most of the abundance changes would be clearly incompatible with
the various $T_{\rm eff}$-log\,$g$ criteria and furthermore would only
introduce other  severe modifications of the stellar abundances,
which would be less easily explainable within the context of a 
born-again giant.
Most of the elements showing an increasing or decreasing abundance
trend are not more sensitive than the elements whose abundance
has remained constant in the same time-span.
The observation of a definite trend in abundance over time 
for all elements with a suspected abundance change, further supports
the finding of a changing chemical composition. 
Likewise, the consistency between the different epochs for most
elements also speaks against the stellar parameters being seriously
in error. 

Many of the spectral lines of the species with changing chemical
content are by necessity relatively strong, which could cause
problems in their interpretation as abundances. 
The effects of hyperfine and isotope splittings have been investigated
for a few Li, Sc, Y and Ba lines but found to be insignificant
except for the Li\,{\sc i}\,D doublet; the Li abundances
in Table \ref{t:abund} have been derived from such synthesis. 
Since the current analysis is based on the LTE approximation,
departures from LTE may introduce spurious results which
could be interpreted as abundance variations. 
We, however, find it unlikely that such departures may be
responsible for all of the observed drastic variations 
(in some cases 1.0\,dex), since 
(1) in most cases the abundances
are derived from lines of the dominant species in the atmosphere 
(exceptions: Li, Zn and Rb),
(2) the line scatter of the relevant elements is similar or
smaller than for the other elements,
(3) in general the abundances are based on more lines than for
the typical elements, 
(4) other elements with similar atomic structures, which thus
should experience similar departures from LTE, do not
show any abundance variations, and 
(5) it seems less likely that the relatively minor 
differences in stellar parameters between the four observing
epochs could introduce {\em differential} NLTE effects
of $\sim 1.0\,$dex.
Naturally though, it cannot be excluded that 
NLTE effects may be accountable for parts
of the observed abundance alterations for some elements,
e.g. for Ti and Cr (see below).

As will be further discussed below, the abundance variations can
be naturally explained as a result of the mixing and 
nucleosynthesis which has occurred
following the final He-shell flash 
-- H-burning, Li-production and $s$-processing --
which furthers supports the conclusion that the changes are real.
The elements which should not participate in any nuclear reactions
to any significant extent by the He-shell flash do not show
any abundance variations, thereby making the whole abundance pattern
consistent with expectations. 

In the absence of any plausible alternative explanation, 
we are thus forced to the conclusion that the observed 
changes in chemical composition are most likely real.
The abundance alterations first noted by Asplund et al. (1997b)
are thus confirmed by the present, more detailed
analysis covering significantly more observational data. 

\subsection{Nucleosynthesis
\label{s:nucleo}}

The current chemical composition of Sakurai's object is
clearly not pristine.
The very evolved nature of the star is evidenced 
by the H-deficiency and high abundances of heavy
elements, not characteristic of neither disk dwarfs
nor halo dwarfs (Edvardsson et al. 1993). 
Sakurai's object seem to have undergone H-burning, 
followed by He-burning and a second phase of CNO-cycling,
as well as associated nuclear reactions such as
Li-production and $s$-processing.

Sakurai's object is slightly metal-poor. With a 
C/He ratio of 10\% the Fe mass fraction is 
0.2\,dex (mean of the four observing epochs)
below solar, but had instead the input carbon
abundance been used the mass fraction would be
0.7\,dex lower as a result of the carbon problem. 
However, only [Mg/Fe]$=-0.1$, [Ca/Fe]$=0.1$, [Ti/Fe]$=0.2$
and [Cr/Fe]$=0.0$ (mean of April, May, June and July for Ti and Cr)
are consistent with such an origin while all other 
elements seem to have been modified throughout
the evolutionary history of Sakurai's object, 
with the possible exceptions of Si ([Si/Fe]$=0.7$)
and S ([S/Fe]$=0.5$); also Ti and Cr seem currently
to be modified (Sect. \ref{s:abundchanges}). 
The most obvious property is, of course, the
presence of H-burning material due to the 
H-deficiency and high He-content.
Furthermore, the high [Na/Fe]$=1.4$ and [Al/Fe]$=1.0$
abundance ratios can probably be explained by proton-captures on
$^{22}$Ne and $^{25}$Mg.
The significant C and O ([O/Fe]$=1.4$) enhancements 
require He-burning
through the 3$\alpha$-process followed by subsequent
$\alpha$-captures. 

The overabundant N ([N/Fe]=2.0) can not be accounted
for by CNO-cycling of the initial CNO nuclei but requires
a second phase of H-burning in which C and O produced by
He-burning are converted to N.
Such a conclusion is supported by the very high 
[Ne/Fe]$=2.4$; Ne has presumably been synthesized from
$\alpha$-captures on products of He-burning and CNO-cycling
(since Mg is not overabundant, Ne has more likely been 
formed from $^{14}$N seeds rather than $^{16}$O nuclei).
Also the low $^{12}$C/$^{13}$C ratio in Sakurai's object
strongly suggests a second stage of CNO-cycling following
He-burning, since the $^{13}$C abundance should be very
low after $\alpha$-processing;
the observed $^{12}$C/$^{13}$C ratio 
($1.5 \leq ^{12}$C$/^{13}$C$ \leq 5$)
encompasses the
equilibrium value of about 3.5 from CNO-cycling.
The high $^{13}$C/N ratio requires though that the
proton supply was exhausted before conversion of the
available C to N 
was completed.
The Si and S abundances may also require nuclear processing,
as in other H-deficient stars (Lambert et al. 1998),
but the channels and conditions under which this would 
have occurred are still unclear. 
The observed Li has most likely been synthesized from
the Cameron-Fowler (1971) Be transport mechanism.

Sakurai's object is strongly enhanced
in $s$-process elements, in particular the light 
$s$-elements. Also the very high abundance ratios  
[K/Fe]=0.8, [Sc/Fe]=1.6 (October), [Ni/Fe]=0.9, [Cu/Fe]=1.8 and 
[Zn/Fe]=1.7 (October) are likely explained 
by neutron captures of lighter seed nuclei (presumably
mainly S, Ar, Ca and Fe), as previously suggested for
R\,CrB stars (Asplund et al. 1998; Lambert et al. 1998)
and FG\,Sge (Gonzalez et al. 1998). 
Very likely $^{13}$C is the neutron source;
the alternative reaction $^{22}$Ne($\alpha$,n)$^{25}$Mg
which ignites at higher temperatures,
seems to be ruled out since the required neutron exposure
($\approx 10$ neutrons per Fe seed nucleus) is not consistent
with the observed Mg/Fe ($\approx 1$) ratio.
All the elements Ni-La
could be synthesized with a {\em single} neutron exposure
$\tau \sim 0.2 \pm 0.1$\,mb$^{-1}$; the Sr/Rb ratio indicates
a neutron density of $N_{\rm n} \sim 10^8$\,cm$^{-3}$
(Malaney 1987). An exponential
exposure of neutrons provides less good agreement.
The fact that also the abundances of Ni, Cu and Zn are well 
explained with a single neutron exposure indicates
that most of the envelope has been exposed to neutrons
and little dilution with gas which has not suffered
$s$-processing has occurred. 
Though [Ba/Fe]=0.6 is not atypical for AGB-stars, many of 
the light $s$-elements have very extreme ratios
(e.g. [Sr/Fe]=3.0 and [Y/Fe]=2.2 in May) not observed in
AGB-stars or post-AGB stars evolving towards the 
planetary nebula-phase for the first time, 
which may suggest that these elements have 
been synthesized following departure from the AGB.
Indeed, both the Li and the (light) $s$-element abundances
seem to have increased dramatically throughout 1996, 
due to mixing and processing which was initiated
in connection with the final He-shell flash.

\section{The evolution of Sakurai's object in 1996}

\subsection{Changes in chemical composition
\label{s:abundchanges}}

Though most elemental abundances remained the same
throughout 1996, a few elements are
distinguished by their changing content within only a
few months. These variations 
may be the result of either mixing of a previously modified
chemical composition or current nuclear processing.
The mixing time-scale is determined by the convective
time-scale which is on the order of a couple of months
for Sakurai's object. 
The nuclear burning time-scale is comparable since
it is determined by convection to bring fresh protons
to H-burning temperatures. Thus, both 
processes may in principle be able to explain the
extremely rapid evolution of Sakurai's object. 

\subsubsection{Mixing processes}

Rather than reflecting current nucleosynthesis in the star, 
the elemental abundance variations may simply be
due to mixing of the surface layers with gas from the interior.
If the initial envelope abundances are significantly different from
those of the interior, a changing chemical 
content may be observed without invoking present
nuclear processing.
Such a depth-dependent chemical composition may either be the
result of previous nucleosynthesis (which must have occurred {\em during}
the AGB-phase), or through dust-gas separation (which must have 
occurred {\em after} departure from the AGB).

Pure mixing of the envelope with the H-depleted gas of the
interior could in principle explain the decreasing 
H-content. However, some amount of H-burning is expected
after the final He-shell flash on theoretical grounds
(Renzini 1979), since the flash should have caused an extensive
convection zone ingesting the still H-rich
envelope, which would bring the gas to H-burning temperatures.
Though mixing may explain part of the diminishing H abundance,
it is not possible to invoke mixing to explain the increasing
Li content. Since Sakurai's object is H-deficient and
H-burning necessarily destroys any present amount of the much
more fragile Li, the observed high Li abundance can not
have been inherited from a previous AGB-phase. 
The observed Li/H ratio ($10^{-6.1} - 10^{-4.8}$) far exceeds 
the value in even the most Li-rich AGB-stars ($\approx 10^{-8}$). 

The observed changes of the light
$s$-elements may be the consequences of mixing with gas
exposed to neutrons in between the thermal
pulses on the AGB now brought to the surface. 
Unfortunately, the exact nature of the 
last flashes immediately before departing from the AGB is 
poorly known, but a preferential production of the light
$s$-elements as observed is not inconceivable. Whether the
very extreme [$s$/Fe] ratios
(e.g. [Rb/Fe]=2.9, [Sr/Fe]=3.3 and [Y/Fe]=2.9 in October) 
could be accomplished remains to be shown, however. As
noted in Sect. \ref{s:nucleo} the abundance distribution
of the heavy elements is better described with a single
neutron exposure rather than an exponential exposure
as expected for thermal pulses in AGB-stars, 
which may indicate that the 
$s$-elements are not inherited from the previous AGB-phase.
The fact that the heavy element abundances in Sakurai's
object even before the observed abundance alterations in 1996
are quite different from those observed in
post-AGB stars evolving towards the planetary nebula-regime
for the first time, also seems to support this conclusion.

The observed abundance increases in Ti and Cr are more
problematic to explain as a result of nucleosynthesis, whether
from a previous phase or presently occurring.
In principle, Ti and Cr could be synthesized through
$s$-processing, as is the case with Sc.
Both Ti and Cr are, however, preceded by elements (Sc and V
respectively) which have significantly lower abundances.
The neutron capture cross-sections (Beer et al. 1992) 
do not suggest a
predominant build-up of Ti and Cr at the expense of the
surrounding elements. 
(Though Zn succeeds the less abundant Cu for October 1996, 
the calculations of Malaney (1987) show that a single neutron 
exposure may produce such an abundance distribution due to
the differences in cross-sections.) 
Furthermore, to explain the high Cr abundance for October,
the Ca abundance would have had to decrease if it were
the seed, which is not observed.
Conceivably, repeated $\alpha$-captures could explain
the Ti and Cr abundance increases, in particular in the
light of the high [Ne/Fe], [Si/Fe] and [S/Fe] ratios, 
but it would seem to require unreasonably
high temperatures. Furthermore, Ca is not significantly enhanced. 
In the light of a lacking plausible nucleosynthetic
explanation, it is not precluded that a combination
of errors (e.g. NLTE effects, unaccountable blends) may have conspired
to produce spurious results, especially since
the interpretation of an abundance alteration of Ti and
Cr is largely controlled by the October spectrum, while
the [Ti/Fe] and [Cr/Fe] ratios for the other dates
are representative of slightly metal-poor stars.
For the moment, we are unable to find a nucleosynthetic
explanation for the observed trends in Ti and Cr.
We note that a similar problem of observed high 
[Ti/Fe] and [Cr/Fe] ratios also exists for some R\,CrB 
stars (Lambert et al. 1998).

Alternatively, mixing may cause variations not because
the interior has undergone previous phases of 
nucleosynthesis, but because the surface layers have
experienced dust fractionation, as e.g. suggested for
post-AGB stars and FG\,Sge 
(Bl\"ocker \& Sch\"onberner 1997). 
Such a suggestion may seem able
to explain the observed increases in Sc, Ti, Cr and Y, since
those elements can be heavily depleted in
post-AGB stars (e.g. Gonzalez et al. 1997). However, the scenario 
is unlikely to be the main solution to the observed variations. 
Elements such as Al, Ca, Fe and Ni should then also show similar
or stronger trends which are not observed, and the increase in
Zn abundance is directly at odds with its inability to condense
into dust (Cardelli 1994). 
Also, the variations in the H and Li content cause severe
problems for the suggestion. 
Furthermore, dust depletion cannot explain the
abundance ratios relative to Fe during the later times,
in particular the extreme $s$-element ratios
(e.g. [Rb/Fe]=2.9, [Sr/Fe]=3.3 and [Y/Fe]=2.9 in October),
when the abundance pattern should be less affected by the
actions of dust fractionation.
It is also hard to understand why exactly those elements
which are likely to be synthesized in connection with a 
final He-shell flash (H, Li, and the $s$-elements)
should be particularly modified while
no other elements show similar trends.
We therefore find such an explanation rather contrived. 
Similar arguments can be constructed against the suggestion
of dust fractionation being responsible for the observed
variations in the abundances of the $s$-elements in
FG\,Sge (Langer et al. 1974; Bl\"ocker \& Sch\"onberner 1997).

\subsubsection{Nuclear processing}

\begin{figure}[t!]
\centerline{
\psfig{figure=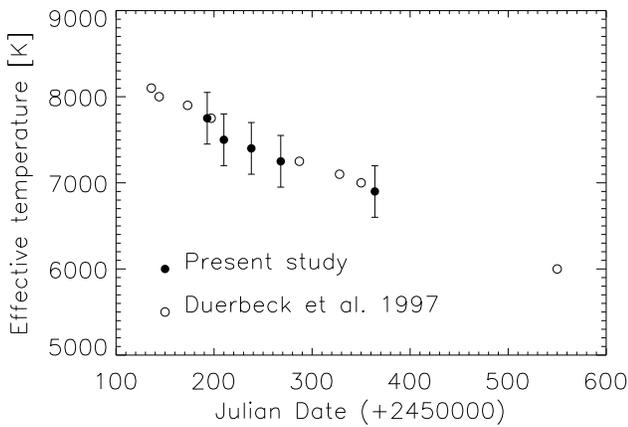,width=9.cm}}
\caption{The evolution of the effective temperature of Sakurai's object 
since discovery in 1996 and 1997 as judged from spectroscopy 
(present study) and $UBVRiz$ photometry (Duerbeck et al. 1997). The estimated
temperatures reveal an essentially linear decrease with time. 
It should be noted
that the last photometric data point is significantly more uncertain
than the other estimates
}
         \label{f:sakteff}
\end{figure}

The most notable abundance alteration in 1996 is the decreasing
H-content by a factor of 10, due to 
the ingestion of the H-rich envelope 
in connection with the He-flash. 
Since most of the luminosity is now probably provided by 
H-burning rather than He-burning (Renzini 1990), the diminishing
H-content is in fact the nuclear fuel of the star being consumed.
Not only is the generated energy sufficient to provide
the high luminosity of the star, the released energy
is also causing the expansion of the star to giant dimensions. 
Assuming a mass of $10^{-4}$\,M$_{\sun}$ of the ingested material and an
initial H content of log\,$\epsilon_{\rm H} = 10.0$ (April),
the liberated energy over six months produces a luminosity
which exceeds the classical Eddington limit by about a factor
of 6 (see also Sect. \ref{s:tefflum}). 
The generated energy during these months is less
than the estimated binding energy of the stellar envelope outside
the He-burning shell (assuming the shell being located at a
stellar radius of
$\sim 2 \cdot 10^9$\,cm and a mass of the convective shell of
$\sim 10^{-2}$\,M$_{\sun}$, Renzini 1990).
However, assuming instead an initial solar
H-abundance prior to the He-shell flash in the ingested material
is sufficient to provide the required energy for lifting the outer layers 
to produce a born-again giant. 
These simple-minded estimates naturally assume that the
diminishing H-content only reflects the effects of H-burning
rather than mixing of the envelope with previously
H-depleted gas from the interior.

The increased Li abundance is most easily explained by
Li-production through the Cameron-Fowler mechanism. 
Since H-burning also destroys present Li, 
the observed Li cannot have been generated in a 
previous AGB-phase but must be currently produced in
the star, simultaneous with the
consumption of H. Furthermore, since the required $^3$He,
which is produced during the main-sequence phase, is
eradicated during He-burning, the Li-production must
proceed in an environment 
not previously exposed to He-burning. 

The high $^{13}$C abundance provides an efficient
neutron source for the $s$-processing 
through $^{13}$C($\alpha,n)^{16}$O.
The abundance alterations of the light $s$-process 
elements (Zn, Rb, Sr, Y and Zr, as well as Sc) can be interpreted
as being due to current synthesis in the star; it is less 
clear whether this is also
true for the heavy $s$-elements though there are indications
that Ba may have increased in 1996. 
As discussed above, the abundance distribution of the heavy
elements suggests that
the variations are not due to pure mixing, but rather to ongoing
$s$-processing.
If this interpretation is correct, however, it may pose
a serious challenge for any theoretical modelling.
The ignition of CN-cycling due to initiated flash-driven
envelope convection at a temperature of $T\la 10^8$\,K
should cause a splitting of the shell into two convective
zones, the lower one burning He at the bottom, 
while the upper experiencing CN-cycling. As described above, Li can
be produced in the upper zone (``hot-bottom-burning''),
but the $^{13}$C produced there does not release neutrons
unless the temperature is significantly higher 
($T \ga 1.5 \cdot 10^8$\,K), 
i.e. in conditions typical of the lower convection zone.
A possible solution may be that H-burning has essentially
run to completion during the peak He-burning phase, in which
case the two convection zones would once again connect, and
thereby bringing down $^{13}$C to the relevant high 
temperatures for neutron release to occur.  
No doubt, the final He-shell flash in Sakurai's object was
a highly dynamical phenomenon with a very short
timescale ($\la $months), which may well have 
been severely inhomogeneous, similar
to what is observed in simulations of supernovae. Such asymmetric
mixing may perhaps explain the simultaneous Li-production
and $s$-processing. 
It is even possible that the surface layers may not have been
chemically homogeneous during 1996 and further variations in
the observed abundances of H, Li, $^{13}$C and
the $s$-elements could be expected.

As discussed in detail above, the increase in Ti and Cr abundances
presents a problem when interpreting the variations
as reflecting nuclear processing.

\subsubsection{Summary}

The most likely explanation for the observed changes
in chemical composition of Sakurai's object is thus 
the direct consequences of the mixing 
{\em and} nuclear reactions which
has occurred following the He-shell flash: ingestion
and burning of the H-rich envelope through (interrupted) CN-cycling, 
Li production from
the simultaneously ingested $^3$He, and $s$-processing
from neutrons liberated through $\alpha$-captures on
the available $^{13}$C, which in turn has 
been produced from the concomitant proton captures on $^{12}$C.
In fact, the abundance alterations are in 
excellent accordance with theoretical expectations (Renzini 1990),
but the exact details how in particular the $s$-processing would
have proceeded is less clear. 
However, we are presently not able to convincingly rule out the
possibility that the $s$-elements have been synthesized in between
previous He-shell flashes on the AGB and are now being brought
to the surface purely through dredge-up.
Such mixing though is not possible to invoke to explain 
the Li-enrichment or all of the diminishing H-content, 
which requires current nucleosynthesis.
In any way, the observed changes can only be understood
if the final He-shell flash occurred {\em after} the H-burning
shell had been extinguished in the post-AGB star, since otherwise
the entropy barrier from H-burning would prevent ingestion 
of the H-rich envelope and no substantial mixing or 
nucleosynthesis besides He-burning would occur.
The observed variations of Ti and Cr are, however, not easily
explained within this framework.

\subsection{Changes in effective temperature and luminosity
\label{s:tefflum}}

\begin{figure}[t!]
\centerline{
\psfig{figure=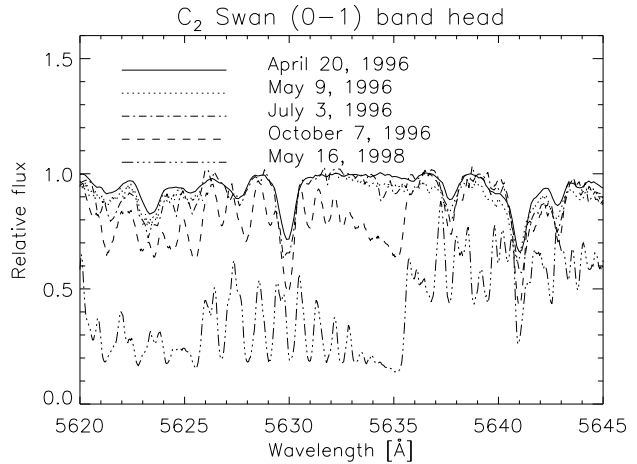,width=9.cm}}
\caption{The evolution of the C$_2$ 0-1 Swan bandhead at
5635\,\AA$ $ in Sakurai's object
during 1996, as well as a recent spectrum acquired in May 1998, 
illustrating the increasing strengths of molecular
features due to the decreasing $T_{\rm eff}$. Note that the
continuum placement for the May 1998 spectrum is highly
uncertain due to the heavy blending 
}
         \label{f:c2}
\end{figure}

\begin{figure}[t!]
\centerline{
\psfig{figure=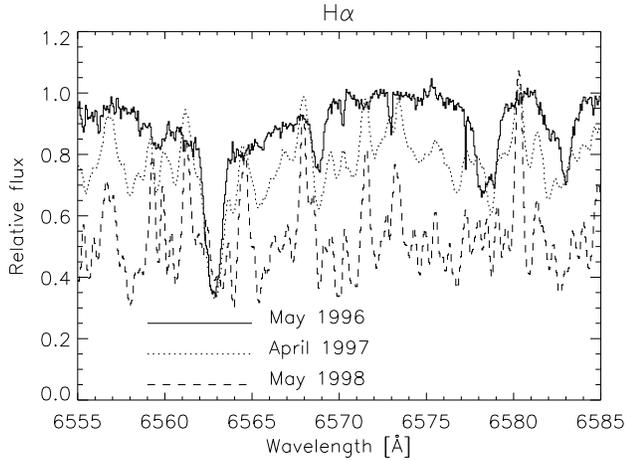,width=9.cm}}
\caption{The evolution of the H$\alpha$ feature in Sakurai's object from
just after discovery (May 1996), to April 1997 and  
May 1998. Note that the
continuum placement for the May 1998 spectrum is highly
uncertain due to the severe line-crowding 
}
         \label{f:h1}
\end{figure}

Sakurai's object has evolved rapidly
in effective temperature following discovery, as evident
both from the spectroscopic analyses presented here for 1996 and
from $UBVRiz$ photometry (Duerbeck et al. 1997) covering 1996 and early
1997, as illustrated in Fig. \ref{f:sakteff}. 
The photometric temperatures have been derived using similar H-deficient
model atmospheres as those applied here, which shows 
the good agreement between spectroscopic
and photometric temperature estimates in 1996.
Comparing the analysed spectra of Sakurai's object in 1996 reveals
clear signatures of a decreasing $T_{\rm eff}$ which was already
noted by Asplund et al. (1997b): from April to October
pronounced C$_2$ and CN bands developed
(Fig. \ref{f:c2}) and the H lines became weaker as apparent from Fig. \ref{f:hbeta} (also due to the diminishing H-content), an evolution
which has continued in 1997 (Kerber et al. 1997) and 1998 (Fig. \ref{f:h1},
though the recent development of heavy line-blanketing in the
visual region makes the weakening of the H-lines less apparent).
In the beginning of 1998 the stellar spectrum resembles a  
carbon star (Figs. \ref{f:c2} and \ref{f:h1}).
The latter does not, however, imply
similarly low $T_{\rm eff}$ as in carbon stars since the H-deficiency
and C-enhancement make molecular features containing C very
prominent (Asplund et al. 1997a). 
Rather $T_{\rm eff}$ is more likely around 5000-6000\,K,
though a detailed abundance analysis of these very crowded spectra
will not be an altogether easy task, as obvious from
Figs. \ref{f:c2} and \ref{f:h1}. 
This rapid cooling is the result of the stellar expansion
caused by the drastically increased energy release after the
final He-shell flash.
A similar phenomenon is well-known to occur in novae
in which the photosphere expands at roughly constant luminosity down
to about 6000\,K, when the expanding shell becomes optically thin.
We would thus expect the luminosity to have
remained roughly constant during the expansion.

From $UBVRiz$ monitoring of Sakurai's object, Duerbeck et al. (1997)
find evidence for a luminosity increase of about 30\%. 
It is, however, not unlikely that this conclusion may be compromised
by adoption of inappropriate bolometric corrections for the star;
considering the H-deficiency and its peculiar composition
the line-blanketing
will be very different compared with normal supergiants of similar
temperatures. A constant luminosity can therefore in our opinion
not be ruled out; clearly more detailed calculations of bolometric
corrections for the star are needed.

\begin{figure}[t!]
\centerline{
\psfig{figure=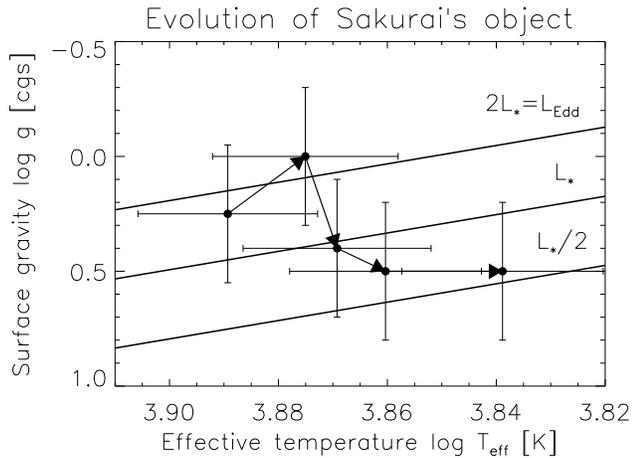,width=9.cm}}
\caption{The spectroscopically derived fundamental parameters of Sakurai's
object in 1996 suggest
an evolution at decreasing stellar luminosity when taken at face
value (but see text). The solid lines illustrate slopes of 
constant luminosity each separated by a factor of 2. The middle
line corresponds to $L_* = 10^{4.4}$\,L$_{\sun}$ when assuming 
$M_* = 0.8\,$M$_{\sun}$. The line denoted by $2L_*$ is identical
to the classical Eddington limit (pure electron scattering for 
fully ionized gas) for H-deficient stars
}
         \label{f:param}
\end{figure}

Our stellar parameters offer information on the luminosity:
L $\propto T^4_{\rm eff}/g$ according to standard relations. 
Fig.\nolinebreak \ref{f:param} shows three loci of constant L in 
the log $g$ versus $T_{\rm eff}$ plane together with
the derived values. 
When taken at face value, the spectroscopic parameters suggest
an evolution with a decreasing luminosity; however, the parameters
are consistent with a constant luminosity and perhaps also with the 
modest increase suggested by Duerbeck et al. (1997)
within the errors of measurement.
Spectroscopic estimates of luminosity are dependent on the assumption
about the stellar mass but the primary uncertainty may arise from
the basic assumptions underlying the construction of the
model atmospheres.
As pointed out by Asplund et al. (1997b), the assumption of hydrostatic
equilibrium for the star may be inappropriate. In particular,
an underestimation of the surface gravity will result from  neglecting the
hydrodynamics due to, for example, an overall atmospheric
expansion or turbulent pressure (Gustafsson et al. 1975).
The assumption of constant luminosity and the observed decrease 
in $T_{\rm eff}$ require e.g. the expansion velocity to have
{\it increased} during 1996. This conclusion follows directly
from the second derivative of the luminosity 
$L = 4\pi R^2 \sigma T_{\rm eff}^4$:

\begin{equation}
0 = 
\frac{2}{T_{\rm eff}}\frac{{\rm d}^2T_{\rm eff}}{{\rm d}t^2} -
\frac{2}{T_{\rm eff}^2}\left(\frac{{\rm d}T_{\rm eff}}{{\rm d}t}\right)^2 +
\frac{1}{R}\frac{{\rm d}^2R}{{\rm d}t^2} -
\frac{1}{R^2}\left(\frac{{\rm d}R}{{\rm d}t}\right)^2.
\end{equation}

\noindent Due to the linear decrease in $T_{\rm eff}$ with time in 1996,
${\rm d}^2T_{\rm eff}/{\rm d}t^2 = 0$ and thus
the expansion velocity must have increased:
${\rm d}^2R/{\rm d}t^2 = {\rm d}v_{\rm exp}/{\rm d}t > 0$.
Duerbeck et al. (1997) arrived at the same conclusion based on
an analysis of photometric data.
The observed trend of a decreasing $\xi_{\rm t}$
also suggests that hydrostatic equilibrium may have been
inapplicable directly following discovery.
Furthermore, for both April and May the spectroscopically
derived parameters are located at the classical Eddington
luminosity for electron scattering (Asplund 1998). 
Thus, we cannot at present rule out an evolution at constant luminosity
from the spectroscopy.

Assuming a stellar mass of 0.8\,M$_{\sun}$ (see below), 
the spectroscopically
derived parameters for October indicate a stellar luminosity of 
$10^{4.1}$\,L$_{\sun}$, which is in very good accordance with
predictions for born-again giants 
(Iben \& MacDonald 1995; Bl\"ocker \& Sch\"onberner 1997).
This corresponds to a distance of about 7\,kpc (using the
above-mentioned $E(B-V)=0.7$),
which agrees with previously estimated distances
of 5.5-8\,kpc (Duerbeck \& Benetti 1996; Duerbeck et al. 1997)
but is in sharp contrast to the short distance of 1.1\,kpc
claimed by Kimeswenger \& Kerber (1998) (see also Jacoby et al. 1998).
Adopting the spectroscopic parameters of the earlier dates
would require a larger distance.
A distance of only 1.1\,kpc would imply log\,$g = 2.1$
for October, a value well outside the acceptable uncertainties
in the spectroscopic analysis.

The very rapid evolution of Sakurai's object following discovery
is significantly faster than for FG\,Sge 
(van Genderen \& Gautschy 1995) though seemingly similar to that
of V605\,Aql though available data is scarce 
(Lundmark 1921; Clayton \& De Marco 1997). FG\,Sge
has throughout the last century brightened visually while
cooling, while V605\,Aql reached peak brightness within
only two years, before starting to show R\,CrB-type
fadings and turning too faint for followed monitoring
(Seitter 1987). Sakurai's object is currently evolving on
a time-scale of a few years; the star probably started
to brighten in 1994 (Duerbeck et al. 1997). 
It is of interest to compare these short evolutionary time-scales
with theoretical predictions. 

According to Renzini (1990) the time-scale for burning the
ingested H-envelope (which generates more 
luminosity than the initial He-burning which triggered
H-burning) after the He-shell flash is 

\begin{equation}
\tau_{\rm H-burn} \approx \frac{E_{\rm H} \Delta M_{\rm H}}{40 L_{\rm pl}},
\end{equation}

\noindent where $E_{\rm H}$ denotes the energy release from H-burning
(CNO-cycling), $\Delta M_{\rm H}$ the amount of ingested H, and
$L_{\rm pl}$ the luminosity plateau during the post-AGB evolution;
the numerical factor takes into account that the H-burning proceeds
at a luminosity roughly 40 times higher than $L_{\rm pl}$.
For a typical case of a 0.6\,M$_{\sun}$ post-AGB star, $\tau_{\rm H-burn}$
will be about 25\,years (Renzini 1990), 
which is in good agreement with the only
published numerical calculation of a final He-shell flash event
by Iben \& MacDonald (1995), in which the subsequent brightening occurred
over 17\,years
Furthermore, the theoretical time-scale is in agreement
with the observed time-scale for H-burning in Sakurai's object
(Sect. \ref{s:abundchanges}).
Different evolutionary time-scales are expected with different 
stellar masses $M_*$, since $\Delta M_{\rm H}$ is a very sensitive function
of $M_*$ (e.g. Bl\"ocker \& Sch\"onberner 1997). 
The very rapid evolution of Sakurai's object and V605\,Aql can
thus presumably be accomplished with a slightly higher $M_*$ and 
therefore smaller $\Delta M_{\rm H}$. 
The main difference in evolutionary time-scales between FG\,Sge on one
hand and Sakurai's object and V605\,Aql on the other 
is therefore probably due to different $M_*$.

The life-time of the quiescent born-again giant phase may be 
about a factor of
40 longer than $\tau_{\rm H-burn}$, since the generated energy
will be radiated away at a luminosity not very different from 
$L_{\rm pl}$ (Renzini 1990). 
In reality, the life-time will be shorter since (1) the
star can be found only part of the time in the giant regime and
(2) due to the effects of mass-loss.
Typical life-time as a bright giant of 100-1\,000\,years can
thus be expected, which is not incompatible with
the observational evidence;
the visual brightness of V605\,Aql has decreased greatly since
outburst in 1919 but the bolometric luminosity has probably remained
roughly the same (Harrison 1996; Clayton \& De Marco 1997). 
Judging from their similarities in rise-time 
and visual variability at peak brightness, it is not a bold 
speculation that Sakurai's object will soon also form an optically
thick surrounding dust cloud, which will cause a 
visual fading it may not recover from;
the development
of an IR excess (Eyres et al. 1998) is certainly suggestive of 
such a prospect.
The very rapid brightening of V605\,Aql (and now Sakurai's object)
may perhaps have endowed a dynamical  
ejection of the outer envelope (Asplund 1998) which subsequently
underwent dust condensation, while the more gentle brightening
of FG\,Sge did not provide the necessary conditions for formation
of a completely obscuring dust shell but only the irregular
dust condensations typical of R\,CrB stars.

\section{Relation to FG\,Sge, V605\,Aql and the R\,CrB stars}

Sakurai's object bears several striking similarities with the
R\,CrB stars (Asplund et al. 1997b; Lambert et al. 1998), in
particular with the R\,CrB star V854\,Cen (Asplund et al. 1998);
from an abundance perspective Sakurai's object can 
be classified as an R\,CrB star
\footnote{Due to the different adopted C/He ratios in 
the analyses of Sakurai's object and the R\,CrB stars, it
is not possible to directly compare the {\it absolute} 
abundances, since those are proportional to the C/He ratio.
Had instead C/He=1\% been used for Sakurai's object, then 
all abundances would have been 1.0\,dex lower to 
first order (Asplund et al. 1998). Therefore it is more
appropriate to compare {\it relative} abundances 
for different stars.}.
Recently, Sakurai's object has also been reported to show
visual variability (Liller et al. 1998): 
in February 1998 the star had faded by almost two magnitudes
which by April it had recovered from,
a behaviour characteristic of the R\,CrB stars.
The irregular variability has since then continued
(H. Duerbeck, private communication).
Thus, Sakurai's object fulfills the two defining properties
of the R\,CrB stars: a distinct H-deficiency and unpredictable
visual dimming events, presumably due to dust formation episodes
in the vicinity of the stars. 
For the first time, the ``birth'' of an R\,CrB star has therefore
been witnessed, proving the ``Final Flash'' scenario (Renzini 1979)
as a viable channel for the formation of R\,CrB stars,
though perhaps not the only option. 

The R\,CrB stars can be put in two groups by chemical composition:
 a homogeneous
majority class and a relatively heterogeneous minority class
(Lambert et al. 1998), which are distinguished by a 
lower metallicity (here: Fe/C) and several extreme abundance ratios, 
in particular [Si/Fe] and [S/Fe]. 
Sakurai's object, however, presently has traits of both
groups. It is, therefore, not possible for the moment to identify
which, if either, of the two R\,CrB groups evolved from a star experiencing a
final flash. 
Sakurai's object lacks the extreme
[Si/Fe] and [S/Fe] ratios of the minority, but has e.g. their
low Fe/C ratio and high [Na/Fe] ratio.
Furthermore, in no minority member has Li previously
been detected, which is a feature in four of the
majority stars (UW\,Cen, R\,CrB, RZ\,Nor and SU\,Tau).
Lithium, however, could conceivably be destroyed as Sakurai's object 
continues to evolve.
Sakurai's object resembles most closely the R\,CrB star
V854\,Cen, whose status as a majority or minority member
is also unclear (Asplund et al. 1998).
The only property of Sakurai's object
 not known to be shared with the R\,CrB stars is the
high $^{13}$C content (Asplund et al. 1997b). Possibly, the high $^{13}$C
abundance is
a transient feature. It should be noted too that even rough estimates of
the $^{13}$C abundance are not available for most R\,CrB stars.

Additional support for the final flash scenario comes from FG\,Sge
and V605\,Aql. 
Currently FG\,Sge is showing typical R\,CrB-like variability 
(Papousek 1992) and there are hints that it may be
H-deficient (Gonzalez et al. 1998).
Unfortunately the published abundance analyses 
(e.g. Kipper \& Kipper 1993; Gonzalez et al. 1998) of FG\,Sge
differ significantly, even in [X/Fe], which prevents
a detailed comparison with Sakurai's object and the R\,CrB stars, 
in particular since a normal H-abundance
has been assumed for the analyses.
FG\,Sge is, however, also significantly enhanced in the heavy
$s$-elements and not only in the lighter $s$-elements, as 
is the case in Sakurai's object. 
During outburst, the spectrum of V605\,Aql resembled closely 
those of cool R\,CrB stars (Lundmark 1921; Ludendorff 1922;
Clayton \& De Marco 1997), though detailed information 
on its chemical composition is lacking.

The situation is therefore slightly unfortunate. All of the
identified born-again giants Sakurai's
object, FG\,Sge and V605\,Aql show similarities with the R\,CrB stars,
clearly suggesting the final flash scenario being a possible route
to forming R\,CrB stars. However, none of the three stars
are readily identified with neither the majority nor the minority
group.
It is tempting to tentatively identify at least 
the minority stars as descendants of a final flash, because of
their more heterogeneous nature and their possibly higher
C/He ratios (Asplund et al. 1997b; Asplund et al. 1998;
Lambert et al. 1998), which are in better agreement with
theoretical predictions (Iben \& MacDonald 1995;
Sch\"onberner 1996). 
By applying  Occam's razor, one could
speculate that {\it all} R\,CrB stars have indeed been
formed through a final He-shell flash in a cooling pre-WD. 
Until the abundance differences between the stars
can be explained either by their earlier nucleosynthetic
evolution or  through the work of dust-gas separation
(Lambert et al. 1998), this conclusion seems premature.
Here we, therefore,  are happy to conclude that at least one R\,CrB star
-- Sakurai's object --
is a final flash object.
Given the striking similarities in terms of chemical composition
between V854\,Cen and Sakurai's object (Asplund et al. 1998) it is
very likely that at least these two stars share the same evolutionary
background.

\section{Conclusions}

An analysis of the available published and  
unpublished spectra of the born-again giant Sakurai's
object from 1996 confirms the extremely rapid evolution of the star
previously noted by Asplund et al. (1997b). Throughout 1996 Sakurai's
object cooled significantly by about 1\,000\,K, which is also obvious
from photometry (Duerbeck et al. 1997);
the cooling has also continued until the present (1998). 
Such a cooling is presumably
the result of the expansion of the photosphere following the drastically
increased luminosity due to the He-shell flash. 
Even more spectacular, the chemical composition of 
Sakurai's object shows definite signs of having been significantly 
altered within only a few months (from April to October 1996).
Again, these changes are most likely interpreted as being caused by
the mixing and nuclear reactions which have ensued as a result of the
final He-shell flash: ingestion and burning of the H-rich envelope,
Li-production through the Cameron-Fowler mechanism, and $s$-processing
of the light $s$-elements (including Sc). 
To our knowledge, Sakurai's object represents the fastest case 
of stellar evolution ever observed when disregarding complete
stellar disruptions such as supernovae.

Sakurai's object shows strong abundance similarities with the R\,CrB stars,
not only a distinct H-deficiency and C-enhancement. 
In all respects, Sakurai's object
would be classified as an R\,CrB star judging from a compositional
perspective (Lambert et al. 1998). 
Since Sakurai's object also recently has started showing irregular
fading episodes typical of the R\,CrB stars (Liller et al. 1998),
Sakurai's object indeed seem to have evolved into
being an R\,CrB star. 
Thus, for the first time ever, the "birth" of an R\,CrB star may have been 
witnessed, which lends strong support for the final flash scenario
as a probable channel for forming at least some of the R\,CrB stars.
The exact relation between Sakurai's object and the majority and
minority classes of the R\,CrB stars needs, however, to be clarified 
before one draws any conclusions on how many of the R\,CrB stars have been
formed through a final He-shell flash in a post-AGB star on the 
WD cooling track.  

Further studies of the future evolution of Sakurai's object is clearly needed. 
Judging from its rapid metamorphosis already since discovery in 1996,
it is not unlikely that the star will provide astronomers with
even more surprises and spectacular changes of appearance in the
years to come. 
In this respect, both a close photometric monitoring of the visual
variability and spectroscopic studies of the changes in chemical
composition after 1996 are of great importance. 
The latter will, however, require a brave soul in order to analyse
the terrifying richness of atomic and molecular lines, many of
which are yet unidentified lines presumably due to $s$-elements. 
As long as the IR excess is not too disturbing (Eyres et al. 1998), 
it may be advantageous to use high-resolution near-IR spectrum 
for such abundance analyses. 
Furthermore, we encourage studies on the chemical composition of
FG\,Sge, in particular regarding its H-content, and the R\,CrB
star U\,Aqr, which shows very pronounced $s$-element enhancements
similar to FG\,Sge and Sakurai's object (Bond et al. 1979),
in order to assess better the relationship
between born-again giants and the R\,CrB stars. 
Finally and most urgently, 
the theoretical modelling of the final He-shell flash events needs to be
extended
to a range of stellar masses and initial conditions with a detailed study of
the mixing processes and 
including the relevant nuclear reactions such as CNO-cycling, 
He-burning, Li-production
and $s$-processing, which will likely provide strong constraints
on the models. 
Admittedly, such modelling is a Herculean task considering the
dynamical nature of the event, but is unfortunately required
in order to fully understand the rapid evolution of Sakurai's object
and related stars.
   
\begin{acknowledgements}
We are grateful to G. Gonzalez, M. Keane, V. Klochkova
and G. Wallerstein for obtaining spectra of Sakurai's object.
The referee H. Duerbeck provided us with useful suggestions.
MA acknowledges financial support connected with a post-doctoral
fellowship at Nordita. 
DLL's research has been supported in part by the U.S.
National Science Foundation (grant NSF 9618414) and the Robert A. Welch
Foundation of Houston, Texas. The financial
support TK has received through the Estonian Science Foundation 
grant 3166 is acknowledged.
\end{acknowledgements}


\end{document}